\documentclass[aps,prd,preprint,a4paper]{revtex4}

\begin{document}

\title{On the rare $B_{s}\rightarrow \mu\mu$ decay and 
noncontractibility of the physical space}

\author{Davor Palle}
\affiliation{
Zavod za teorijsku fiziku, Institut Rugjer Bo\v skovi\' c \\
Bijeni\v cka cesta 54, 10000 Zagreb, Croatia}

\date{April 16, 2012}

\begin{abstract}
It is very well known that the rare electroweak processes could be
very sensitive to the physics beyond the Standard Model.
These processes are described with quantum loop diagrams 
containing also heavy particles. We show that the electroweak
theory with the noncontractible space, as a symmetry-breaking
mechanism without the Higgs scalar, essentially changes 
the Standard Model prediction of
the branching ratio of the $B_{s}$ meson decaying to two muons.
The branching ratio is lower by
more than 30\% compared with the Standard Model result.
Although the measurements are very challenging, the implications
on the selection of the symmetry-breaking mechanism could be 
decisive. 
\end{abstract}

\maketitle 

\section{Introduction and motivation}

In the previous paper \cite{PalleCP}, we show that the CP
violation in the K and B meson systems is affected by the
short distance electroweak (EW) corrections of the BY theory \cite{Palle1}
that differ substantially from the Standard Model (SM) ones.
We also show that the QCD in noncontractible space (BY is the ultraviolet nonsingular
theory) enhances the standard QCD t-quark charge asymmetry and the effect is
observed at the Tevatron \cite{Palle2}.
It is natural to expect the deviation from the SM predictions
for the rare electroweak processes which are described by the similar
loop diagrams.

In this paper, we concentrate on the evaluation of the short 
distance EW corrections in the SM and the BY theory
to the decay $B_{s}\rightarrow \mu\mu$. In the next chapter, we
present the SM and BY results for the effective transition 
operator consisting of the Z-boson penguin and box diagram
contributions \cite{Inami}. The last chapter is devoted
to the numerical evaluations and discussion.

\section{$B_{s}\rightarrow \mu\mu$ amplitude} 

The SM calculation of the $B_{s}\rightarrow \mu\mu$ decay
can be found in Ref.\cite{Inami} and the QCD correction
in Ref.\cite{Buchalla}.

The branching ratio for the $B_{s}\rightarrow \mu\mu$ decay
is equal to \cite{Buchalla}:

\begin{eqnarray}
Br(B_{s}\rightarrow \mu\mu)&=&\tau(B_{s})\frac{G_{F}^{2}}{\pi}
\left(\frac{\alpha_{e}}{4\pi\sin^{2}\Theta_{W}}\right)^{2}
F_{B_{s}}^{2}m_{\mu}^{2}m_{B_{s}}\sqrt{1-4\frac{m_{\mu}^{2}}{m_{B_{s}}^{2}}}
  \nonumber \\
&\times & \mid V_{ts}^{*}V_{tb} \mid^{2}Y^{2}(x_{t}),  
\end{eqnarray}
\begin{eqnarray*}
x_{t}\equiv (\frac{m_{t}}{M_{W}})^{2},\ 
Y(x_{t})=\eta_{Y}Y_{0}(x_{t}),\ \eta_{Y}=1.012=QCD\ correction\ factor,\ 
 \\
\langle 0 \mid (\bar{s}b)_{V-A,\nu} \mid B_{s}(p)\rangle = \imath
F_{B_{s}}p_{\nu},\ 
Y_{0}(x)=\frac{x}{8}(\frac{4-x}{1-x}+\frac{3x}{(1-x)^{2}}\ln x).
\end{eqnarray*}

Thus, the short distance part is hidden in the gauge
invariant term $Y(x_{t})$ \cite{Inami,Buchalla}.
We must perform the calculation within the BY theory 
\cite{PalleCP,Palle1}. We choose the 't Hooft-Feynman
gauge to make direct comparison with the SM results \cite{Inami}.
Although the Nambu-Goldstone scalars do not couple to
Dirac fermions in the BY theory, we can use them as 
the auxiliary fields in our calculus. The gauge
invariant observables could be evaluated in the unitary
gauge without the Nambu-Goldstone scalars. 

The new ingredient of the BY theory
is the UV cutoff in the spacelike domain
of the Minkowski spacetime \cite{Palle1}:
$\Lambda=\frac{\hbar}{c d}=\frac{2}{g}\frac{\pi}{\sqrt{6}}
M_{W} \simeq 326 GeV$.
$\Lambda$ is fixed by the Wick's theorem and the trace anomaly with
the Lorentz and gauge invariant weak coupling and the weak gauge boson
mass. 

Because of the conservation of the electromagnetic current
of the muon pair, there is no $\gamma$ penguin contribution to
$B_{s}\rightarrow \mu\mu$. The Z boson penguin ($\Gamma_{Z}$) and the box
(C) diagram contribution (see Figs.(1-4) and Appendix of 
Ref.\cite{Inami}) have the following expressions 
(with the abbreviation $\sin \Theta_{W}=s_{W}$):

\begin{eqnarray}
Y_{0}(x_{j}) = \frac{1}{2}(\Gamma_{Z}+C)(x_{j})-(x_{j}\rightarrow
x_{u}),\ j=c,t,\  \Gamma_{Z}\equiv \sum^{h}_{i=a}
\Gamma^{(i)},  
\end{eqnarray}
\begin{eqnarray}
b\bar{s}Z\ vertex =\Gamma_{Z\mu}^{(i)}=\frac{1}{(4\pi)^{2}}
\frac{g^{3}}{\cos \Theta_{W}} V^{*}_{js}V_{jb}\bar{s}\gamma_{\mu}
P_{L}b \Gamma^{(i)},  
\end{eqnarray}
\begin{eqnarray*}
\Gamma^{(a+b)}&=&(-\frac{1}{2}+\frac{1}{3}s_{W}^{2})(1+\frac{1}{2}
\frac{m_{j}^{2}}{M_{W}^{2}})B_{1}(0;m_{j},M_{W}), \hspace{50 mm} \\
\Gamma^{(c)}&=& -\frac{1}{4}(-1+\frac{4}{3}s_{W}^{2})
(B_{0}(0;m_{j},M_{W})+m_{j}^{2}L(m_{j},M_{W}))+
\frac{2}{3}s_{W}^{2}m_{j}^{2}L(m_{j},M_{W}),  \\
\Gamma^{(d)}&=& -\frac{1}{2}\frac{m_{j}^{2}}{M_{W}^{2}}
(\frac{1}{3}s_{W}^{2}B_{0}(0;m_{j},M_{W})
+m_{j}^{2}(\frac{1}{2}-\frac{1}{3}s_{W}^{2})L(m_{j},M_{W})),  \\
\Gamma^{(e)}&=& -\frac{3}{2}(1-s_{W}^{2})(B_{0}(0;m_{j},M_{W})
+M_{W}^{2}L(M_{W},m_{j})),  \\
\Gamma^{(f+g)}&=& -s_{W}^{2}m_{j}^{2}L(M_{W},m_{j}),  \\
\Gamma^{(h)}&=&\frac{1}{8}(-1+2s_{W}^{2})\frac{m_{j}^{2}}{M_{W}^{2}}
(B_{0}(0;M_{W},m_{j})+M_{W}^{2}L(M_{W},m_{j})),  
\end{eqnarray*}
\begin{eqnarray}
C=\frac{1}{2}M_{W}^{2}L(M_{W},m_{j}),\ 
x_{j}\equiv (\frac{m_{j}}{M_{W}})^{2}.
\end{eqnarray}

The external fermion masses and momenta are neglected, while 
neutrinos are assumed to be massless.
We can compare our expressions with those of Inami and Lim, term
by term, acknowledging the following Green functions (only real
parts are considered, see \cite{Palle1}):

\begin{eqnarray*}
\frac{\imath}{16\pi^{2}}B_{0}(k^{2};m_{1},m_{2})&\equiv &
(2\pi)^{-4}\int d^{4}q(q^{2}-m_{1}^{2})^{-1}((q+k)^{2}-m_{2}^{2})^{-1},
  \hspace{30 mm}\\
\frac{\imath}{16\pi^{2}}k_{\mu}B_{1}(k^{2};m_{1},m_{2})&\equiv &
(2\pi)^{-4}\int d^{4}q q_{\mu}(q^{2}-m_{1}^{2})^{-1}
((q+k)^{2}-m_{2}^{2})^{-1},  \\
\frac{\imath}{16\pi^{2}}L(m_{1},m_{2})&\equiv &
(2\pi)^{-4}\int d^{4}q(q^{2}-m_{1}^{2})^{-2}(q^{2}-m_{2}^{2})^{-1},
 \\
B_{1}(0;m_{1},m_{2})&=&\frac{1}{2}(-B_{0}(0,m_{1},m_{2})
+(m_{2}^{2}-m_{1}^{2})\frac{\partial B_{0}}{\partial k^{2}}
(0,m_{1},m_{2})),  \\
B_{0}(0;m_{j},M_{W})&=&\Delta_{UV}-x_{j}\ln x_{j}/(x_{j}-1),\  
\Delta_{UV}\equiv UV\ infinity,  \\
\frac{\partial B_{0}}{\partial k^{2}}(0;m_{1},m_{2})
&=&\frac{1}{2}\frac{m_{1}^{2}+m_{2}^{2}}{(m_{1}^{2}-m_{2}^{2})^{2}}
-\frac{m_{1}^{2}m_{2}^{2}}{(m_{2}^{2}-m_{1}^{2})^{3}}
\ln \frac{m_{2}^{2}}{m_{1}^{2}}, \\
m_{j}^{2}L(m_{j},M_{W})&=& x_{j}(x_{j}-1)^{-2}(1-x_{j}+\ln x_{j}).
\end{eqnarray*}

Adding up all graphs, we get:

\begin{eqnarray}
\Gamma_{Z}(x_{j})&=&\frac{1}{48}(x_{j}-1)^{-2}\{-[(x_{j}-1)(66+9 x_{j}
-15 x_{j}^{2}+2 s_{W}^{2}(-34+29 x_{j}+5x_{j}^{2}))] \nonumber \\
&+& 12 x_{j}(2+3 x_{j})\ln x_{j} \} , 
\end{eqnarray}
\begin{eqnarray}
C(x_{j})= \frac{1}{2}(\frac{1}{x_{j}-1}-\frac{x_{j}}{(x_{j}-1)^{2}}
\ln x_{j}).
\end{eqnarray}

It seems that Inami and Lim \cite{Inami} make an approximation
for the total Z-boson penguin contribution. Our expressions
for any single graph (a-h) coincide with the Inami-Lim expressions $(A\cdot 1)$
of Ref.\cite{Inami}. Their total Z-boson contribution is:

\begin{eqnarray}
\Gamma_{Z}(x_{j})_{approx}=\frac{1}{4}x_{j}-\frac{5}{4}
\frac{1}{x_{j}-1}+\frac{1}{4}\frac{3 x_{j}^{2}+2 x_{j}}{(x_{j}-1)^{2}}
\ln x_{j}.
\end{eqnarray}

It is easy to get necessary Green functions with the UV cutoff
of the BY theory from the preceding integral representations:

\begin{eqnarray*}
B_{0}^{\Lambda}(0;m_{1},m_{2})&=&\int^{\Lambda^{2}}_{0}
dy \frac{y}{(y+m_{1}^{2})(y+m_{2}^{2})} \\
&=& (m_{1}^{2}\ln\frac{\Lambda^{2}
+m_{1}^{2}}{m_{1}^{2}}-m_{2}^{2}\ln\frac{\Lambda^{2}+m_{2}^{2}}
{m_{2}^{2}})/(m_{1}^{2}-m_{2}^{2}), 
\end{eqnarray*}
\begin{eqnarray*}
\frac{\partial B_{0}^{\Lambda}}{\partial k^{2}}(0;m_{1},m_{2})&=&
\frac{1}{2}(\frac{\partial \tilde{B}_{0}^{\Lambda}}
{\partial k^{2}}(0;m_{1},m_{2})+
\frac{1}{2}(\frac{\partial \tilde{B}_{0}^{\Lambda}}
{\partial k^{2}}(0;m_{2},m_{1})),\hspace{90 mm} 
\end{eqnarray*}
\begin{eqnarray*}
\frac{\partial \tilde{B}_{0}^{\Lambda}}
{\partial k^{2}}(0;m_{1},m_{2})&=&
m_{2}^{2}\int^{\Lambda^{2}}_{0}dy \frac{y}{(y+m_{1}^{2})
(y+m_{2}^{2})^{3}} \hspace{100 mm}  \hspace{50 mm} \\
&=& [\Lambda^{2}(-2 m_{1}^{2}m_{2}^{4}+2 m_{2}^{6}+
\Lambda^{2}(m_{2}^{4}-m_{1}^{4}))+
2 m_{1}^{2}m_{2}^{2}(\Lambda^{2}+m_{2}^{2})^{2} \\
&\times & (\ln \frac{m_{1}^{2}}{m_{2}^{2}}-\ln \frac{\Lambda^{2}+
m_{1}^{2}}{\Lambda^{2}+m_{2}^{2}})]
/[2 (\Lambda^{2}+m_{2}^{2})^{2}(m_{2}^{2}-m_{1}^{2})^{3}], 
\end{eqnarray*}
\begin{eqnarray*}
L^{\Lambda}(m_{1},m_{2})&=& -\int^{\Lambda^{2}}_{0}
dy y (y+m_{1}^{2})^{-2}(y+m_{2}^{2})^{-1} \hspace{90 mm} \\
&=& -[\Lambda^{2}(m_{1}^{2}-m_{2}^{2})+m_{2}^{2}(\Lambda^{2}+
m_{1}^{2})(\ln \frac{m_{2}^{2}}{m_{1}^{2}}+
\ln \frac{\Lambda^{2}+m_{1}^{2}}{\Lambda^{2}+m_{2}^{2}})] \\
&\times &
[(\Lambda^{2}+m_{1}^{2})(m_{1}^{2}-m_{2}^{2})^{2}]^{-1}.
\end{eqnarray*}

Now we have all ingredients to make numerical estimates and
comparisons between SM and BY short distance parts of the rare
$B_{s}\rightarrow \mu\mu$ decay.

\section{Results and discussion}

With insertion of the Green functions, with and without
the UV cutoff, into the expressions for the amplitude, one 
can compare branching ratios.
We can safely ignore c-quark contribution in the amplitude since
$\ \ \mid V^{*}_{cb} V_{cs} Y(x_{c}) \mid  
/ \mid V^{*}_{tb} V_{ts} Y(x_{t}) \mid \simeq {\cal O}(10^{-4})$.
Let us numerically inspect the difference between our exact and 
Inami-Lim approximate sum for the Z-boson penguin (Eqs.(5) and (7)):

\begin{eqnarray*}
parameters:\ s_{W}^{2}=0.23,\ m_{u}=3 MeV,\ m_{c}=1.3 GeV,\ 
m_{t}=172 GeV,\\ M_{W}=80.4 GeV,\ \Lambda=326 GeV \  
\rightarrow \Gamma_{Z}(x_{t})_{approx}/\Gamma_{Z}(x_{t})=0.962.
\end{eqnarray*}

We see that the difference is only a few percents. 

Since the
QCD correction to the $Y_{0}$ is evaluated at the scale $\mu 
\simeq m_{b}$ much smaller than $\Lambda$, it does not deviate
from the SM in the BY theory.
The large uncertainty in the quark mixing angles and the meson
form factor $f_{B_{s}}$ is the consequence of nonperturbative QCD hadron
physics. Now we make our final estimate and comparisons for the
branching ratio of the $B_{s}\rightarrow \mu\mu$ decay:

\begin{eqnarray*}
\frac{Br(B_{s}\rightarrow \mu\mu)(\Lambda)}
{Br(B_{s}\rightarrow \mu\mu)(\infty)}=\frac{Y^{2}_{0}(x_{t})\mid_{\Lambda}}
{Y^{2}_{0}(x_{t})\mid_{\infty}}=0.664, \\
\frac{Br(B_{s}\rightarrow \mu\mu)(\Lambda)}
{Br(B_{s}\rightarrow \mu\mu)(\infty)_{approx}}=0.708.
\end{eqnarray*}

The estimate with the UV cutoff (noncontractible space as a
symmetry breaking mechanism) gives a more than $30\%$ lower
branching ratio compared to the SM. For a UV cutoff that is
larger by one order of magnitude, 
the difference diminishes
$Br(10 \Lambda=3260 GeV)/Br(\Lambda=\infty)=0.996$.

The cumulative error of the SM prediction
$Br(SM)=(3.2\pm 0.2)\cdot 10^{-9}$ is below $10\%$ \cite{Buras}, thus
giving the opportunity to the Tevatron, LHCb, SuperKEKB and SuperB to discriminate 
between the two predictions in the very near future \cite{LHCb}.

\end{document}